\documentstyle[aps,pre]{revtex}
\title{Phenomenological approach to non-linear Langevin equations.}

\author{Josep Bonet Avalos and Ignacio Pagonabarraga\\
Departament de F\'{\i}sica Fonamental, Facultat de  F\'{\i}sica,
Universitat
de Barcelona. \\
Diagonal 647, E-08028 Barcelona (Spain). }

\begin{document}

\maketitle

\begin{abstract}
In this paper we address the problem of consistently construct
Langevin equations to describe fluctuations in non-linear systems.  Detailed
balance severely restricts the choice of the random force, but we prove that
this property together with the macroscopic knowledge of the system is not
enough to determine all the properties of the random force. If the cause of the
fluctuations is {\em weakly} coupled to the fluctuating variable, then the
statistical properties of the random force can be completely specified. For
variables odd under time-reversal, microscopic reversibility and weak coupling
impose symmetry relations on the variable-dependent Onsager coefficients.
We then analyze the fluctuations in two cases:  Brownian motion in position
space and an asymmetric diode, for which the analysis based in the master
equation approach is known.  We find that, to the order of validity of the
Langevin equation proposed here, the phenomenological theory is in agreement
with the results predicted by more microscopic models.
\end{abstract}

\parskip 2ex
\renewcommand{\theequation}{\arabic{section}.\arabic{equation}}

\section{Introduction}

\setcounter{equation}{0}

	A widely used method to study fluctuations is the so-called
Langevin approach, introduced by P.  Langevin\cite{Lang} as a way to
study Brownian motion\cite{Ors,vK}.  In his treatment, the interaction
force between the Brownian particle and the bath particles is split
into two contributions.  The first one corresponds to the damping due
to the frictional force exerted by the bath.  The second one arises
from the thermal motion of the bath molecules.  The Brownian particle
then experiences a large number of collisions per unit of time with
the bath molecules, which give rise to a rapidly varying contribution,
responsible for the observed erratic motion of the particle.  Then,
the equation for the velocity of a Brownian particle of unit mass can
be written as

\begin{equation}
\dot{u}(t) = -\xi u(t) + F(t)                         \label{1.1}
\end{equation}

\noindent The first term of the right hand side is the friction force.
This force is a linear function of the variable $u$, $\xi$ being the
friction coefficient.  The second term stands for the Langevin {\em
complementary} force or random force, whose nature is unknown a
priori.  Beyond Langevin's original treatment and with the aim of
determining $u(t)$, several assumptions have been made on the nature of
the random force.  The most basic one is that $F(t)$,
being an unpredictable rapidly varying function of time, is taken
as a stochastic process, often assumed to be Gaussian
and white with zero mean\cite{Ors}. The second moment of $F(t)$ is determined
by means of the fluctuation-dissipation theorem, which relates the strength of
the random
force to the dissipation, i.e.  the friction coefficient, $\xi$.  This
crucial point ensures that the law of equipartition of energy
is satisfied by the system under the effect of $F(t)$.
With all these assumptions, the statistical properties of the random
variable $u(t)$ can be obtained\cite{Ors}.

	Let us consider here a generic macroscopic variable, $X(t)$,
whose relaxation is phenomenologically described by the equation

\begin{equation}
\dot{X}(t) = -A(X(t))                   \label{1.2}
\end{equation}

\noindent where $A(X(t))$ is the flux of the variable which can be a
non-linear function of $X(t)$.  In this paper, we will be interested
in the dynamics of the fluctuations of that variable. The
fluctuating counterpart of $X(t)$ will be denoted by $x(t)$, the former being
obtained from the latter by some averaging procedure.
It has been then proposed that $x(t)$
satisfies the equation

\begin{equation}
\dot{x}(t) = -A(x(t)) + F(t)            \label{1.3}
\end{equation}

\noindent obtained by replacing $X(t)$ by $x(t)$ in the phenomenological
equation (\ref{1.2}) and adding a Langevin force, assumed to be
Gaussian and white.  This approach has been used for systems in which
$A(x)$ is a linear function of the variable $x$\cite{Lan,Ons,dGr}.
However, when $A(x)$ is non-linear, the addition of a random force
$F(t)$ to eq.  (\ref{1.2}), as a procedure to describe the random
process $x(t)$, has often been questioned.  We want to emphasize two
main criticisms\cite{vK2}:

\begin{enumerate}
\item The Gaussian nature of the random process $F(t)$ is
postulated\cite{Ors}, but a more general character of the process
could in
principle be equally plausible.
\item Averaging eq. (\ref{1.3}) with respect to the realizations of
the $F(t)$
given an initial condition $x_0$, identifying $X(t)$ with $\langle
x(t)\rangle_{x_0}$ and comparing with the phenomenological equation
(\ref{1.2}),
one finds that

\begin{equation}
\left\langle A(x(t))\right\rangle_{x_0} \neq A(X(t))    \label{1.4}
\end{equation}

\noindent if $A(x)$ is not a linear function of $x$.  It is then
argued that this difference arises from the fact that the use of
$A(x)$ also in the equation for the fluctuations (\ref{1.3}) is not
justified because $A(x)$ is phenomenological, that is, it is known
except a function of the order of the strength of the
fluctuations\footnote{See also the comments in ref.\cite{Maz1} about
this point.}.  \end{enumerate}

\noindent In view of these criticisms, it is often concluded that the
Langevin approach is not valid for the case of internal (thermal)
noise in non-linear systems and that a more microscopic description is
necessary to derive the proper equation for the
fluctuations\cite{vK3}.

	Recently, however, some attention has been payed to the
properties of the Langevin random force in the case of thermal noise
for a physical system bearing a non-linear $A(x)$.  Let us write
$A(x)$ under the form

\begin{equation}
A(x) \equiv -\beta(x) \frac{d}{dx} \ln P_e(x)             \label{1.5}
\end{equation}

\noindent where $P_e(x)$ is the equilibrium probability density of the
variable $x$, supposed known, and $\beta(x)$ is a given function of
$x$.  Eq.  (\ref{1.5}) emphasizes the fact that the macroscopic flux
is a function of the thermodynamic force

\begin{equation}
{\cal F}(x) = \frac{\partial}{\partial x} \ln P_e(x)      \label{1.5b}
\end{equation}

\noindent such that the flux vanishes when the thermodynamic force also
vanishes. Near equilibrium, $\beta$ tends to a constant and $A(x)$ takes
the form of a linear function of the thermodynamic force ${\cal F}$, in
the spirit of Onsager's theory of irreversible processes. Thus, $\beta(x)$
is the corresponding
variable-dependent Onsager coefficient. P.  Mazur
and D. Bedeaux have proved that
the underlying microscopic reversibility when applied to the Langevin
equation, severely restricts the properties of the Langevin
complementary force, contrary to what is stated in
the first remark.  Indeed, if $x(t)$ in eq. (\ref{1.3}) is a variable {\em even
or odd
under time-reversal} and if $F(t)$ is {\em independent of the variable} $x(t)$,
it is found that\cite{Maz1}:  1) $F(t)$ must be a Gaussian process and 2)
$\beta(x)$ must be a constant.  Note that if $P_e(x)$ is not Gaussian,
then $A(x)$ is non-linear even if $\beta$ is a constant and however,
$F(t)$ still must be a Gaussian process.  Furthermore, another conclusion
that can be drawn from this result is that if $\beta(x)$ is not a
constant, $F(t)$ cannot be independent of $x(t)$. This case will be referred to
as {\em genuinely} non-linear.

	A simple way to construct a random force depending on the
variable $x(t)$ is to assume that\cite{Maz2,Maz2b}

\begin{equation}
F(t) = \tilde{c}(x(t)) \tilde{L}(t)                \label{1.6}
\end{equation}

\noindent often referred to as multiplicative noise.  Here, $\tilde{L}(t)$ is a
stationary white process, independent of $x(t)$
and $\tilde{c}(x)$ is a given non-vanishing function of $x$.

	In general, when proposing a Langevin equation to study thermal
fluctuations in a given system, one would wish
that the properties of the random force would follow from the
phenomenological knowledge of the system.  It is clear, as claimed in
ref.\cite{vK}, that if a detailed microscopic description
is available, there are ways to derive the equation
for the fluctuations of a given variable without ambiguity.  However, we are
interested
in the situations where this information is lacking, although it will be
assumed that there exists an underlying microscopic
description in terms of Hamilton or Schr\"{o}dinger equations, and
that a relation like eq.  (\ref{1.2}) is known.  The aim of this paper
is to use the information at hand to consistently determine a Langevin
equation for $x(t)$ in the genuinely non-linear case.  To
this end and in the same
spirit as in refs.\cite{Maz1}-\cite{Maz2b}, we will use generally valid
properties
such as microscopic reversibility to fix the nature of the random
force.  Furthermore, we will discuss the range of applicability of the
theory developed. To that purpose, we will consider that the
equation
for the evolution of $x(t)$ is given again by the phenomenological law
(\ref{1.2}) plus two
additional contributions. The first one is a variable-dependent random force,
$\tilde{F}(x(t),t)$, defined by

\begin{equation}
\tilde{F}(x(t),t) \equiv \int dx \delta(x-x(t)) \tilde{L}(x,t)
        \label{1.8}
\end{equation}

\noindent where $\tilde{L}(x,t)$ is a stationary and white random process,
depending on a parameter $x$ but
independent of random process $x(t)$, which is characterized by its
cumulants\cite{vK}

\begin{eqnarray}
<\tilde{L}(x,t)>_c &=& 0             \label{2.2a} \\
<\tilde{L}(x_1,t_1) \tilde{L}(x_2,t_2)>_c &=&
\tilde{c}_2(x_1,x_2)\delta(t_1-t_2)
        \label{2.2b} \\
<\tilde{L}(x_1,t_1) \tilde{L}(x_2,t_2) \dots \tilde{L}(x_n,t_n)>_c &=&
\tilde{c}_n(x_1,x_2,\dots,x_n) \delta(t_n-t_1)
\delta(t_n-t_2)
 \dots \delta(t_n-t_{n-1}) \nonumber \\
 & &        \label{2.2c}
\end{eqnarray}

\noindent Note that while $L(x,t)$ is independent of the process $x(t)$, the
statistical properties of $\tilde{F}(x(t),t)$ will depend on the evolution
equation for $x(t)$.

        The second
contribution is
a function $\tilde{\Gamma}(x)$, standing for a possible
modification of the phenomenological flux $A(x)$ due to fluctuations. The
proposed Langevin equation then reads

\begin{equation}
\dot{x}(t) = -A(x(t))+\tilde{\Gamma}(x(t))+ \tilde{F}(x(t),t)
        \label{1.8b}
\end{equation}

\noindent A priori,
both $\tilde{\Gamma}(x)$ and the set of functions $\tilde{c}_i(\{x_i\})$ are
undetermined.
We will specialize in the simplest, although very important, case in which
the random process $L(x,t)$ depends only on one single function $c(x)$,
all the cumulats present in eqs. (\ref{2.2a}-\ref{2.2c}) being related with
$c(x)$. This case will be later identified with the widely used multiplicative
noise\cite{Maz2,Maz2b}. The general case will be analyzed
elsewhere. The presence of
the unknown function $\tilde{\Gamma}(x)$ in eq. (\ref{1.8b}) as well as the
treatment to be developed are the main
differences with respect to the analysis given in refs.\cite{Maz1,Maz2,Maz2b},
permiting a larger degree of freedom with important consequences.

	The paper is organized as follows.  In the next section we
will obtain the master equation for the one-dimensional process
$x(t)$ described by eq.  (\ref{1.8b}) with multiplicative noise. In section 3
we apply detailed
balance to the master equation to show that $\tilde{L}(x,t)$ needs to be
Gaussian, which constitutes an alternative derivation of the results
of ref.\cite{Maz2b}. In addition,
we will also obtain a differential equation involving
$\tilde{c}(x)$ and $\tilde{\Gamma}(x)$, as well as the symmetries imposed to
these functions, reminiscent of the Onsager symmetry relations.  In our
context, such an equation serves to
fix neither $\tilde{c}(x)$ nor $\tilde{\Gamma}(x)$ but it is a mere
relationship between them.  This suggests that detailed balance
alone is
not enough to univocally determine the properties of the additional terms
in eq.  (\ref{1.8b}).  In section 4, we will analyze Brownian
motion in position space to illustrate the main ideas exposed so far.
This particular example will serve us to ilustrate that if the mechanisms
causing the fluctuations are weakly coupled to the fluctuating variable, then
we can univocally determine $\tilde{c}(x)$ and $\tilde{\Gamma}(x)$.  This will
be called the {\em weak coupling} assumption. In the same section,
we also apply these ideas to Alkemade's diode\cite{Alk,vKNL}, often used to
show the failure of the phenomenological treatment of the
fluctuations.  This last example will permit us to analyze the
phenomenological theory developed along the paper to the light of a
more microscopic description and, therefore, determine its range of
applicability.  Section 5 is devoted to the conclusions and to a brief
summary of the ideas sketched in this paper.

\section{The Master Equation}

\setcounter{equation}{0}

	Let us assume that the dynamics of the one-dimensional
variable $x(t)$, {\em even or odd} under time-reversal, is described
phenomenologically, that is, in the absence of fluctuations, by eq.
(\ref{1.2}).  Note that this statement constitutes a definition of the
phenomenological equation.  For the fluctuations we have the Langevin equation
given in eq. (\ref{1.8b}) which we rewrite as

\begin{equation}
\dot{x}(t) =  -\tilde{B}(x(t)) +\tilde{F}((x(t),t)
\label{2.1}
\end{equation}

\noindent with

\begin{equation}
\tilde{B}(x) \equiv -\beta(x) \frac{\partial}{\partial x}\ln P_e(x) -
\tilde{\Gamma}(x)
                \label{2.1b}
\end{equation}

\noindent where the form (\ref{1.5}) for $A(x)$ is used. Note that the
"renormalized" flux $\tilde{B}(x)$ is not completely determined since it
depends on the unknown $\tilde{\Gamma}(x)$. We will define an {\em acceptable}
$\tilde{B}(x)$ as such that, after fixing $\tilde{\Gamma}(x)$, still retains
some dependence on the macroscopically relevant Onsager coefficient $\beta(x)$.

        The random force $\tilde{F}(x(t),t)$, as it stands in eq. (\ref{1.8}),
cannot be univocally interpreted unless a rule to
compute averages of the form $<g(x(t))\tilde{L}(x,t)>$, $g(x)$ being an
arbitrary function, is provided\cite{vK}.  In the literature, this ambiguity is
often
referred to as the It\^o-Stratonovich dilemma. For causality reasons, we
demand\cite{Maz1,Maz2} that the state of the system at $t$ is not correlated
with the random force at the same time, that is

\begin{equation}
<\tilde{F}(x(t),t)> = \int dx <\delta(x-x(t))> <\tilde{L}(x,t)> = 0
        \label{1.6b}
\end{equation}

\noindent which in fact prescribes It\^o's rule. However, from the
mathematical point of view, the
It\^o-Stratonovich dilemma is immaterial in our case since
$\tilde{\Gamma}(x)$, and the properties of $\tilde{L}(x,t)$ are a priori {\em
unknowns} that will be
determined later on, on the basis of physical arguments and in agreement with
the chosen interpretation rule.

        For simplicity's sake, in the calculation which follows we will assume
Stratonovich's rule. Then, for clarity, we introduce $\Gamma(x)$, $F(x(t),t)$,
$L(x,t)$ and the cumulants $\{c_n(x_1,\dots,x_n)\}$, as the quantities
consistent
with Stratonovich's rule while $\tilde{\Gamma}(x)$, $\tilde{F}(x(t),t)$,
$\tilde{L}(x,t)$ and $\{\tilde{c}_n(x_1,\dots,x_n)\}$ will be those consistent
with It\^o's rule. Since both
describe the same physical situation, we must have that the equations

\begin{eqnarray}
\dot{x}(t) &=& \left. -A(x(t))+\Gamma(x) + F(x(t),t)
\right|_{\mbox{\footnotesize Stratonovich \normalsize}}
\label{2.2d} \\
\dot{x}(t) &=& \left. -A(x(t))+\tilde{\Gamma}(x)
+ \tilde{F}(x(t),t) \right|_{\mbox{\footnotesize It\^o \normalsize}}
        \label{2.2e}
\end{eqnarray}

\noindent are equivalent.

        Assuming thus Stratonovich's rule, we introduce the multiplicative
noise as the particular case in which the set of cumulants in eqs.
(\ref{2.2a}-\ref{2.2c}) takes the form

\begin{equation}
c_n(x_1,x_2, \dots, x_n) = \alpha_n c(x_1) c(x_2) \dots c(x_n)  \label{nova1}
\end{equation}

\noindent $\alpha_n$ being constants. This particular choice of the random
process $L(x,t)$ makes it equivalent to $c(x)L(t)$ as in eq. (\ref{1.6}), with
$L(t)$ a stationary
white random process whose cumulants are the set of constants $\{\alpha_i\}$.
Note that the choice done in eq. (\ref{nova1}) does not imply
a Gaussian nature of the random process $L(x,t)$ which, in turn, is going to be
Gaussian if $\alpha_i =0$ for $i \geq 3$.

	To obtain the master equation for the probability, we will
essentially follow ref.\cite{Maz1}.  Let us first consider the density
distribution

\begin{equation}
\rho(x,t) \equiv \delta(x-x(t))            \label{2.3}
\end{equation}

\noindent where $x(t)$ is the solution of eq.(\ref{2.2d}) for a
particular realization of $L(x,t)$ and a given initial condition $x_0
\equiv x(t=0)$.  Then, the conditional probability, $P(x,t|x_0)$, for
the random variable $x(t)$ to be at the point $x$ at $t$, given that
it was at the point $x_0$ at $t=0$, is related to $\rho(x,t)$ by

\begin{equation}
P(x,t|x_0) = <\rho(x,t)>_{x_0}             \label{2.4}
\end{equation}

\noindent where, again, the averages are over all the realizations of
$L(x,t)$ and the subindex $x_0$ indicates that the same initial
condition $x_0$ is considered.  Clearly,

\begin{equation}
P(x,t=0|x_0) = \delta(x-x_0)            \label{2.5}
\end{equation}

        In appendix A we show that $P(x,t|x_0)$ satisfies

\begin{equation}
\frac{\partial}{\partial t} P(x,t|x_0)= \left[\frac{\partial}{\partial
x} B(x)
+ \sum_{n=2}^{\infty} (-1)^n \frac{1}{n!} \alpha_n \left(
\frac{\partial}{\partial x} c(x) \right)^n \right] P(x,t|x_0)
\label{2.14}
\end{equation}

\noindent which is the Kramers-Moyal expansion of the master
equation\cite{vK}

\begin{equation}
\frac{\partial}{\partial t} P(x,t|x_0) = \int dx' \left[w(x'
\rightarrow x)
P(x',t|x_0)  -w(x \rightarrow x') P(x,t|x_0)\right] \equiv \int dx' \,
\hat{\cal
W}(x|x') P(x',t|x_0)            \label{2.15}
\end{equation}

\noindent Here, $w(x \rightarrow x')$ stands for the transition
probability from the state $x$ to $x'$, and the operator $\hat{\cal
W}(x|x')$ is defined as

\begin{equation}
\hat{\cal W}(x|x') \equiv w(x' \rightarrow x) - \delta(x-x') \int dx''
\, w(x
\rightarrow x'')                     \label{2.16}
\end{equation}

\noindent Eq. (\ref{2.14}) can be expressed in a more compact form by
introducing the operator $\hat{\cal L}(x)$:

\begin{equation}
\hat{\cal L}(x) \equiv \frac{\partial}{\partial x} B(x)
+ \sum_{n=2}^{\infty} (-1)^n \frac{1}{n!} \alpha_n \left(
\frac{\partial}{\partial x} c(x) \right)^n      \label{2.17}
\end{equation}

\noindent Then

\begin{equation}
\frac{\partial}{\partial t} P(x,t|x_0)= \hat{\cal L}(x) P(x,t|x_0)
                                \label{2.18}
\end{equation}

\noindent We thus identify

\begin{equation}
\hat{\cal W}(x|x') = \hat{\cal L}(x) \, \delta(x-x')    \label{2.19}
\end{equation}

        Furthermore, due to the Markovian nature of the process under
discussion, we can write

\begin{equation}
P(x,t) = \int dx' \, P(x,t|x',t') \, P(x',t')   \label{2.20}
\end{equation}

\noindent where $P(x,t)$ is the one-point probability density.
Setting $x'=x_0$ and $t'=0$ in this last equation, multiplying both
sides of eq.  (\ref{2.18}) by $P(x_0,0)$ and integrating over $x_0$
one gets the evolution equation for $P(x,t)$

\begin{equation}
\frac{\partial}{\partial t} P(x,t)= \hat{\cal L}(x) P(x,t)
\label{2.21}
\end{equation}

\section{Detailed Balance}

\setcounter{equation}{0}

	A consequence of the time-reversibility of the microscopic
equations of motion in a closed system in equilibrium is the so-called
{\em detailed balance}\cite{dGr2}.  When the dynamics of the variable
$x(t)$ is described by a master equation like eq.  (\ref{2.15}),
detailed balance is the condition imposing that for each pair of
states $x$ and $x'$ the transitions must balance:

\begin{equation}
w(x' \rightarrow x) P_e(x') = w(\epsilon x \rightarrow \epsilon x')
P_e(\epsilon x)        \label{3.1}
\end{equation}

\noindent with $\epsilon = 1$ for even and $\epsilon = -1$ for odd variables
under time-reversal.
Furthermore, the equilibrium probability distribution satisfies
$P_e(x)=P_e(\epsilon x)$.

        Detailed balance as shown in eq. (\ref{3.1}) leads to a similar
relationship for the operator $\hat{\cal W}(x|x')$ defined in the previous
section (eq. (\ref{2.19}))\cite{vK}

\begin{equation}
\hat{\cal W}(x|x') P_e(x') = \hat{\cal W}(\epsilon x'|\epsilon x) P_e(x)
\label{3.1b}
\end{equation}

\noindent which, in turn, leads to the relationship

\begin{equation}
\hat{\cal L}(x) \, P_e(x) \, \psi(x) = P_e(x) \hat{\cal
L}^{\dagger}(\epsilon x) \psi(x)
                \label{3.2}
\end{equation}

\noindent where $\psi(x)$ is an arbitrary function. The operator
$\hat{\cal L}^{\dagger}(x)$ is the adjoint of $\hat{\cal L}(x)$ in the
usual sense and is given by

\begin{equation}
\hat{\cal L}^{\dagger}(x) \equiv -B(x) \frac{\partial}{\partial x} +
\sum_{n=2}^{\infty} \frac{1}{n!} \alpha_n \left(c(x)
\frac{\partial}{\partial
x} \right)^n            \label{3.3}
\end{equation}

\subsection*{Even variables}

        Let us first discuss the case of even variables and thus replace
$\epsilon$ by $1$ in eq. (\ref{3.2}). Making use now of eqs. (\ref{2.17}) and
(\ref{3.3}) in eq. (\ref{3.2}) we get

\begin{eqnarray}
\lefteqn{\left\{\frac{\partial}{\partial x} B(x)
+ \sum_{n=2}^{\infty} (-1)^n \frac{1}{n!} \alpha_n \left(
\frac{\partial}{\partial x} c(x) \right)^n \right\} \, P_e(x) \psi(x)
}
\nonumber \\
& & = P_e(x) \,
\left\{-B(x) \frac{\partial}{\partial x} +
\sum_{n=2}^{\infty} \frac{1}{n!} \alpha_n \left(c(x)
\frac{\partial}{\partial
x} \right)^n \right\} \psi(x)           \label{3.4}
\end{eqnarray}

\noindent We now multiply both sides of this last equation by $c(x)$ obtaining

\begin{equation}
\left\{ {\cal D} B(x)
+ \sum_{n=2}^{\infty} (-1)^n \frac{1}{n!} \alpha_n
{\cal D}^n c(x)\right\} \, P_e(x) \psi(x)=  P_e(x) \left\{ -B(x)
{\cal D} +
\sum_{n=2}^{\infty} \frac{1}{n!} \alpha_n c(x) {\cal D}^n \right\} \psi(x)
                        \label{3.5}
\end{equation}

\noindent where the we have introduced the operator ${\cal D}$ as

\begin{equation}
{\cal D} \equiv c(x) \frac{\partial}{\partial x}        \label{3.6}
\end{equation}

\noindent Expressing the left hand side of eq.  (\ref{3.5}) in terms
of the independent operators ${\cal D}^n$, one has

\begin{eqnarray}
\left[ {\cal D} B(x)P_e(x) \right]\psi(x) &+& B(x)P_e(x) {\cal D}
\psi(x)
+ \sum_{n=2}^{\infty} (-1)^n \frac{1}{n!} \alpha_n
\sum_{m=0}^{n} \left( \begin{array}{c} n \\ m \end{array} \right)
\left[ {\cal
D}^{n-m}  c(x) P_e(x) \right] {\cal D}^m \psi(x) \nonumber \\
 &=&  P_e(x) \left\{ -B(x)
{\cal D} +
\sum_{n=2}^{\infty} \frac{1}{n!} \alpha_n c(x) {\cal D}^n \right\} \psi(x)
                \label{3.7}
\end{eqnarray}

\noindent Since $\psi(x)$ is arbitrary, we can now equate the
coefficients of the terms proportional to ${\cal D}^m \psi(x)$ in both sides of
this last equation.  From
the coefficient of ${\cal D}^0 \psi(x)=\psi(x)$, we obtain

\begin{equation}
\left[ {\cal D} B(x)P_e(x) \right] + \sum_{n=2}^{\infty} (-1)^n
\frac{1}{n!}
\alpha_n  \left[ {\cal
D}^{n}  c(x) P_e(x) \right] = 0    \label{3.8}
\end{equation}

\noindent Dividing this equation by $c(x)$ and using eq. (\ref{3.6}),
one gets the stationarity condition $\hat{\cal L}(x) P_e(x)
=0$ as follows from eq.  (\ref{2.21}).  Similarly, from the coeffcient
of ${\cal D} \psi(x)$ we get

\begin{equation}
B(x)P_e(x) + \sum_{n=2}^{\infty} (-1)^n \frac{1}{n!}
\alpha_n \, n \left[ {\cal
D}^{n-1}  c(x) P_e(x) \right] = - B(x) P_e(x)    \label{3.9}
\end{equation}

\noindent Finally, the coefficient of ${\cal D}^m \psi(x)$ for $m \geq
2$ gives

\begin{equation}
\sum_{n=m}^{\infty} \frac{1}{n!} (-1)^n \alpha_n
\left(\begin{array}{c} n \\ m
\end{array} \right) \left[{\cal D}^{n-m} c(x) P_e(x) \right] =
\frac{1}{m!}
\alpha_m c(x) P_e(x); \mbox{\hspace{0.5cm} for $m \geq 2$}
\label{3.10}
\end{equation}

\noindent This last expression stands for an homogeneous system of
equations for the set of constants ${\alpha_i}$ for $i \geq 2$ whose
coefficients are functions of $x$.  Note that $\alpha_2$ can only
appear in the equation for $m=2$ in view of the lower limit of the sum on
the left hand side of eq.  (\ref{3.10}).  However, for $m=2$ the term
containing $\alpha_2$ in the sum
is exactly canceled by the term on the right hand side, so that
$\alpha_2$ is not determined by this system of equations. In
appendix B we proof that the only acceptable solution of eqs. (\ref{3.10}) is
$\alpha_i=0$ for $i\geq 3$.  Therefore, $L(x,t)$ in eq.  (\ref{2.2d}) must be
Gaussian. This important result was already found in ref.\cite{Maz2b} by using
a different procedure.

	Thus, eqs.  (\ref{3.8}) and (\ref{3.9}) now take respectively the
form

\begin{eqnarray}
\frac{\partial}{\partial x} \left\{ B(x) P_e(x) +  \frac{1}{2}
\alpha_2  \left[ c(x) \frac{\partial}{\partial x}  c(x) P_e(x)
\right]\right\}
&=& 0                           \label{3.11} \\
2 B(x) P_e(x) + \alpha_2  \left[c(x) \frac{\partial}{\partial x}  c(x)
P_e(x)
\right] &=& 0                        \label{3.12}
\end{eqnarray}

\noindent according to the definition of ${\cal D}$.
Notice that eq.  (\ref{3.11}) can be obtained from eq.
(\ref{3.12}) by differentiating with respect to $x$.  While the former
is the stationarity condition which states that the divergence of the
probability flux must be zero in equilibrium, the latter says that the
probability flux itself must also vanish.

\subsection*{Odd variables}

        The treatment of the case of odd variables cannot be carried out
following the same procedure as done for even variables for a general
non-vanishing $c(x)$. It is, however, of particular importance the situation in
which $c(x)$ verifies $c(x)=c(-x)$ since we will prove that $L(x,t)$ is
Gaussian
{\em if and only if} $c(x)$ is an even function of its argument\footnote{This
was implicitly assumed in eq. (2.7) of ref.\cite{Maz2b}.}.
Furthermore, we will also determine the corresponding symmetry properties of
$B(x)$ and discuss the consequences that arise from these facts.

        Let us replace $\epsilon$ by $-1$ in eq. (\ref{3.2}) and assume that
$c(x)$ is even. We then get

\begin{eqnarray}
\lefteqn{\left\{\frac{\partial}{\partial x} B(x)
+ \sum_{n=2}^{\infty} (-1)^n \frac{1}{n!} \alpha_n \left(
\frac{\partial}{\partial x} c(x) \right)^n \right\} \, P_e(x) \psi(x)
}
\nonumber \\
& & = P_e(x) \,
\left\{B(-x) \frac{\partial}{\partial x} +
\sum_{n=2}^{\infty} (-1)^n \frac{1}{n!} \alpha_n \left(c(x)
\frac{\partial}{\partial
x} \right)^n \right\} \psi(x)           \label{3.17}
\end{eqnarray}

\noindent Using again the operator ${\cal D}$ and equating the coefficients of
the functions ${\cal D}^m \psi(x)$ at both sides of eq. (\ref{3.17}), we arrive
at

\begin{equation}
\left[ {\cal D} B(x)P_e(x) \right] + \sum_{n=2}^{\infty} (-1)^n
\frac{1}{n!}
\alpha_n  \left[ {\cal
D}^{n}  c(x) P_e(x) \right] = 0    \label{3.18}
\end{equation}

\noindent for the coefficient of ${\cal D}^0 \psi(x)$. From the term
proportional to ${\cal D} \psi(x)$ we get

\begin{equation}
B(x)P_e(x) + \sum_{n=2}^{\infty} (-1)^n \frac{1}{n!}
\alpha_n \, n \left[ {\cal
D}^{n-1}  c(x) P_e(x) \right] =  B(-x) P_e(x)    \label{3.19}
\end{equation}

\noindent The coefficient of ${\cal D}^m \psi(x)$ for $m \geq
2$ gives

\begin{equation}
\sum_{n=m}^{\infty} \frac{1}{n!} (-1)^n \alpha_n
\left(\begin{array}{c} n \\ m
\end{array} \right) \left[{\cal D}^{n-m} c(x) P_e(x) \right] =
(-1)^m \frac{1}{m!}
\alpha_m c(x) P_e(x); \mbox{\hspace{0.5cm} for $m \geq 2$}
\label{3.20}
\end{equation}

\noindent In appendix B we also proof that the only acceptable solution of this
last equation is $\alpha_i =0$
for $i \geq 3$, so that $L(x,t)$ is again a Gaussian process.

        The reverse implication is not difficult to proof. Let us assume that
$L(x,t)$ is a Gaussian process or, what is the same, that $\alpha_i=0$ for $i
\geq 3$. From eq. (\ref{3.2}) with $\epsilon=-1$, making use of eqs.
(\ref{2.17}) and (\ref{3.3}) we obtain

\begin{eqnarray}
\lefteqn{\left\{\frac{\partial}{\partial x} B(x)
+ \frac{1}{2} \alpha_2 \left(
\frac{\partial}{\partial x} c(x) \right)^2 \right\} \, P_e(x) \psi(x)
}
\nonumber \\
& & = P_e(x) \,
\left\{B(-x) \frac{\partial}{\partial x} +
\frac{1}{2} \alpha_2 \left(c(-x)
\frac{\partial}{\partial
x} \right)^2 \right\} \psi(x)           \label{3.20b}
\end{eqnarray}

\noindent where now $c(x)$ is a general non-vanishing function. Equating the
coefficients of the terms proportional to $\partial^2 \psi(x)/\partial x^2$ we
get

\begin{equation}
\frac{1}{2} \alpha_2 \, c^2(x) \, P_e(x) = \frac{1}{2} \alpha_2 \, c^2(-x) \,
P_e(x)          \label{3.20c}
\end{equation}

\noindent and since $c(x)$ is assumed non-vanishing, then $c(x)=c(-x)$.
Therefore,
if detailed balance is satisfied and $x(t)$ is an odd variable under
time-reversal, $L(x,t)$ is
Gaussian if and only if $c(x)$ is an even function.

        Making now use of the Gaussian nature of $L(x,t)$, eqs. (\ref{3.18})
and (\ref{3.19}) become

\begin{eqnarray}
\frac{\partial}{\partial x} \left\{ B(x) P_e(x) +  \frac{1}{2}
\alpha_2  \left[ c(x) \frac{\partial}{\partial x}  c(x) P_e(x)
\right]\right\}
&=& 0                           \label{3.21} \\
(B(x)-B(-x)) P_e(x) + \alpha_2  \left[c(x) \frac{\partial}{\partial x}  c(x)
P_e(x)
\right] &=& 0                        \label{3.22}
\end{eqnarray}

\noindent Again, eq. (\ref{3.21}) has the structure of the divergence of a
flux equal to zero. Since this flux itself is zero
in equilibrium, we get

\begin{equation}
B(x)=-B(-x)                             \label{3.23}
\end{equation}

\noindent Thus, under these conditions eqs. (\ref{3.21}) and
(\ref{3.22}) reduce to eqs. (\ref{3.11})
and (\ref{3.12}) and, therefore, under these circumstances the subsequent
discussion will be valid for even as well as for odd variables.

        While eq. (\ref{3.23}) follows from detailed balance, we will
prove in the next section that if weak coupling is satisfied, this
previous symmetry implies in turn

\begin{eqnarray}
\Gamma(x) &=& -\Gamma(-x) \label{3.24} \\
\beta(x) &=& \beta(-x)  \label{3.25}
\end{eqnarray}

\subsection*{The Fokker-Planck equation}

        Let us consider again eq. (\ref{3.12}). In view of eq. (\ref{2.1b}) it
can be written as

\begin{equation}
\left(\frac{\alpha_2}{2} c^2(x)-\beta(x) \right) \frac{d}{d x}
\ln P_e(x) = \Gamma(x) - \frac{1}{2} \frac{d}{d x} \frac{\alpha_2}{2}
c^2(x)          \label{3.13}
\end{equation}

\noindent Note that the factor $\alpha_2/2$ may be absorbed into
$c^2(x)$.  In
what follows, we will, without lost of generality, choose the scale of
$L(x,t)$ such that $\alpha_2 = 2$. Then, eq. (\ref{3.13}) turns into

\begin{equation}
\left(c^2(x)-\beta(x) \right) \frac{d}{d x}
\ln P_e(x) = \Gamma(x) - \frac{1}{2} \frac{d}{d x}c^2(x)          \label{3.14}
\end{equation}

        With the aim of discussing the original Langevin eq. (\ref{1.8b}) or,
equivalently eq. (\ref{2.2e}),
we have to relate the previous results with those corresponding to \^Ito's
prescription. It is easy to show\cite{vK}, although we will not do it
here, that for $\tilde{\Gamma}(x)$, $\tilde{c}(x)$ and $\tilde{L}(x,t)$,
one can find that i) $\tilde{L}(x,t)$ is also Gaussian with
$\tilde{c}_2(x,x')=\tilde{c}(x)\tilde{c}(x')$, ii) $\tilde{\Gamma}(x)$
and $\tilde{c}(x)$ are related by the equation

\begin{equation}
\left(\tilde{c}^2(x)-\beta(x) \right) \frac{d}{d x}
\ln P_e(x) = \tilde{\Gamma}(x) - \frac{d}{d x}\tilde{c}^2(x)     \label{3.15b}
\end{equation}

\noindent and, for odd variables, iii) the same symmetries as in eqs.
(\ref{3.23}-\ref{3.25}) hold.

Eq. (\ref{3.15}), bearing two unknown functions
$\tilde{\Gamma}(x)$ and $\tilde{c}(x)$,
indicates that for our Langevin equation,
microscopic reversibility alone cannot univocally determine the
properties of the random force $\tilde{F}(x(t),t)$.  Rather, it relates the
functions $\tilde{\Gamma}(x)$ and $\tilde{c}(x)$ with the thermodynamic force
and the
phenomenological coefficient $\beta(x)$. In the next section we will introduce
a plausible way to determine these two functions based on physical grounds.

	Eq. (\ref{3.15b}), the fact that the random
process $\tilde{L}(x,t)$ must be Gaussian as well as the symmetries of
$\tilde{\Gamma}(x)$, $\tilde{c}(x)$ and $\beta(x)$ for odd
variables, constitute the main results of this
section. Some consequences follow.

        Firstly, in view of the fact that
$\tilde{L}(x,t)$
must be a Gaussian process, then eq.  (\ref{1.8b}) is equivalent to a
Fokker-Planck equation of the form

\begin{equation}
\frac{\partial}{\partial t} P(x,t) = \frac{\partial}{\partial x}
\left(
\tilde{B}(x) \, P(x,t) +  \frac{\partial}{\partial x}
\tilde{c}^2(x)
\, P(x,t)
\right)
\label{3.15}
\end{equation}

\noindent Secondly, integrating eq. (\ref{3.15b}) we get

\begin{equation}
\tilde{c}^2(x) = \frac{\Omega}{P_e(x)}+\beta(x) - \frac{1}{P_e(x)}
\int dx' \, P_e(x') \left\{
\frac{d}{dx'} \beta(x') -  \tilde{\Gamma}(x') \right\} \equiv
\beta(x) - \Lambda (x)         \label{4.5}
\end{equation}

\noindent where $\Omega$ is an integration constant.  In this last
expression we have gathered in $\Lambda(x)$ all the terms that
explicitly depend on the equilibrium probability distribution
$P_e(x)$.  Note that, in general, the properties of the random force
are then dependent on the Onsager coefficient $\beta(x)$ as well as on the
equilibrium probability distribution.

        Inserting this last expression for $\tilde{c}^2(x)$ into eq.
(\ref{3.15}), after some algebra, one gets

\begin{equation}
\frac{\partial}{\partial t} P(x,t) = \frac{\partial}{\partial x}
\left(
\beta(x)- \Lambda(x) \right) \left[-P(x,t)
\,\frac{\partial}{\partial x}\ln P_e(x) + \frac{\partial}{\partial x} P(x,t)
\right]		 \label{4.5b}
\end{equation}

\noindent Note that this equation depends on the still unknown
function $\tilde{\Gamma}(x)$ in view of eq.  (\ref{4.5}).  Nevertheless,
independently of that function, the equilibrium probability
distribution $P_e(x)$ is always a solution of this equation.

        Finally, after remark 2 and due to the fact that $\tilde{\Gamma}(x)$
is not necessarily zero for a linear $A(x)$, we have no objective
reason at this point to consider the theory for linear $A(x)$ as
substantially different from that of the genuinely non-linear case.

\section{Two examples}

\setcounter{equation}{0}

	In this section we will analyze two non-trivial examples for
which a more detailed analysis exists.  The first one is the diffusion
of a Brownian particle with a position-dependent friction
coefficient in a general potential field.  The second one is a diode
with a non-symmetric non-linear characteristic function $I=I(V)$

\subsection*{Diffusion of Brownian particles}

	Here, we will study the case in which the variable $x(t)$ is
the position of a Brownian particle in a one-dimensional system
in equilibrium with a reservoir that keeps
constant the temperature $T$.  The particle is sensitive to a
potential field $V(x)$.  Thus, the equilibrium distribution function
is given by

\begin{equation}
P_e(x)=\frac{1}{\cal N} \, e^{-V(x)/kT}         \label{4.1}
\end{equation}

\noindent where ${\cal N}$ is the normalization constant and $k$ is
Boltzmann's
constant. The phenomenological
equation in this case is the relationship between the velocity of the
particle and the force acting on it

\begin{equation}
\dot{X}(t) = -\frac{1}{\xi(X)} \frac{d}{dX} V(X) \equiv -A(X)
\label{4.2}
\end{equation}

\noindent $\xi(X)$ being the friction coefficient which is assumed to
be position-dependent.  Physical examples could be either motion in an
inhomogeneous medium, or an analogy with the three-dimensional
situation where hydrodynamic interactions with other particles or with
the walls of the container causes the friction coefficient of a
Brownian particle to change from one point to another.  From these two
equations we can rewrite $A(X)$ as

\begin{equation}
A(X) = \frac{1}{\xi(X)} \frac{d}{dX} V(X) = -\frac{kT}{\xi(X)}
\frac{d}{dX} \ln
P_e(X)          \label{4.3}
\end{equation}

\noindent which permits us to identify $\beta(X) = kT/\xi(X)$.  To
describe the fluctuations, let us replace $X$ by $x$ and rewrite an
equation of the form of eq.  (\ref{2.1})

\begin{equation}
\dot{x}(t) = \left[\frac{kT}{\xi(x(t))} \frac{d}{dx} \ln
P_e(x)+\tilde{\Gamma}(x(t))
\right]+ \tilde{F}(x(t),t)   \label{4.3b}
\end{equation}

\noindent where the term between brackets is $\tilde{B}(x(t))$.  If detailed
balance has to be satisfied, the corresponding Fokker-Planck equation then
reads

\begin{equation}
\frac{\partial}{\partial t} P(x,t) = \frac{\partial}{\partial x}
\left(
\frac{kT}{\xi(x)}- \Lambda(x) \right) \left[P(x,t)
\,\frac{\partial}{\partial x}
\frac{V(x)}{kT} + \frac{\partial}{\partial x} P(x,t)
\right]		 \label{4.6}
\end{equation}

\noindent Note that this equation does not coincide with the well known
Smoluchowski equation
that describes the diffusion of Brownian particles with
position-dependent friction coefficient (see, for instance,
refs.\cite{vK,Sa}) unless $\Lambda(x)$ identically vanishes.

	At this point we will further assume that {\em the properties
of the random force need to be independent of $P_e(x)$} (This
statement will be referred to as "weak coupling" from now on).  If
this is the case, from eq.  (\ref{3.15b}) we get that, in general,

\begin{eqnarray}
\tilde{c}^2(x)& = &\beta(x)				\label{4.7b}
\\
\tilde{\Gamma}(x) &= &\frac{d}{d x} \beta(x)
\label{4.8b}
\end{eqnarray}

\noindent so that the random force is completely characterized.
To get a more intuitive picture of the physical nature of
this statement, in the case of Brownian motion, we may say that weak
coupling is equivalent to state that the random force should be a
property of the bath system and then it must be independent of the
potential force, $-dV(x)/dx$, externally applied to the Brownian
particle.  Then, if eq.  (\ref{4.8b}) is satisfied, $\Lambda(x)$
identically vanishes and the usual diffusion equation is recovered.

        Note that in the case of odd variables, since we have assumed that
$\tilde{c}(x)$ is even, eqs. (\ref{4.7b}) and (\ref{4.8b}) leads to the
symmetry properties shown in eqs. (\ref{3.24}) and (\ref{3.25}).

	We can then conclude that
our Langevin equation (\ref{1.8b}) is completely determined when detailed
balance and weak coupling are imposed, leading us to the Gaussian
nature of $L(x,t)$ and to eqs.  (\ref{4.7b}) and (\ref{4.8b}).  We have
also found that such a Langevin equation is equivalent to the
Fokker-Planck equation

\begin{equation}
\frac{\partial}{\partial t} P(x,t) = \frac{\partial}{\partial x}
\beta(x) \left[P(x,t) \,\left(-\frac{\partial}{\partial x} \ln P_e(x)
\right)
 + \frac{\partial}{\partial x} P(x,t)
\right]		 \label{4.8c}
\end{equation}

\noindent for the probability distribution.

\subsection*{Diode}

	We will next consider the problem proposed in
refs.\cite{Alk,vKNL}, where an extensive analysis on more microscopic
grounds can be found.  One is now interested in the study of charge
fluctuations in a condenser of capacity $C$ in parallel with a vacuum
diode which is in thermal equilibrium with a reservoir at a
temperature $T$.  The diode is constructed by facing two electrodes
of metals with different work functions\footnote{The work functions
are defined as the work needed to extract an electron from the metal} for the
electrons,
$W_1$ and $W_2$ with $W_1 > W_2$.  The system is
described in the scheme of fig.1.  It is important to note that we
have explicitly included in the figure the contact potential barrier, equal to
the
difference in electrochemical potential per unit charge, when one
electron passes from one metal to the other.  This contact potential is
$\Delta W/e$, where $\Delta W=W_1-W_2$ and $e$ is electron's
elementary charge.  For the system of the figure we have the $I(V)$
characteristics

\begin{equation}
I(V) = I_0 \left( e^{eV/kT}-1 \right)
\label{4.9}
\end{equation}

\noindent The relation between the intensity and the voltage
difference $v$ between the diode plates is obtained by using in eq.
(\ref{4.9}) the fact that $V=\Delta W/e + v$ as it follows from the
figure.  It then reads

\begin{equation}
I(v)  = I_0 \left( e^{\Delta W/kT} e^{ev/kT}-1 \right)   \label{4.9b}
\end{equation}

	One way to proceed is to connect a condenser of
capacity $C$ in parallel with the system of the figure 1 to study
charge fluctuations in the condenser originated in the diode.  In that
case, the equilibrium distribution for the charge in the condenser
follows by standard equilibrium statistical mechanics.  Effectively,
the energy of the condenser when storing a charge $q$ is purely
electrostatic and equal to $E = q^2/2C$.  Then, the equilibrium
distribution is given by

\begin{equation}
P_e(q) \sim e^{-\frac{q^2}{2kTC}}
\label{4.10}
\end{equation}

\noindent With the equilibrium distribution (\ref{4.10}) and the
phenomenological law (\ref{4.9}), one can obtain the equation for the
fluctuations following the same procedure as in the previous example.
The system studied in refs.\cite{Alk,vKNL} is, however, slightly different and
corresponds to that shown in fig.2.  Note that since the system has no
contacts, the plates of the condenser {\em need to be of different
metals}.  This crucial point leads us to the equilibrium distribution
function for the charge in the condenser.  One considers again the
energy of the condenser when a charge $q$ is stored.  To discharge the
condenser, on one hand, the electrons loose electrostatic energy in
traveling from metal $2$ (where the voltage is $0$ according to fig.2)
to metal $1$ (where the voltage is $v=q/C$).  On the other hand, there
is an additional lost of energy due to the fact that the chemical
potential of the electrons in metal $2$ is higher (lower work
function) than in metal $1$ (higher work function).  The energy stored
in the condenser is thus $E = q^2/2C + \Delta W q/e$ and the
equilibrium distribution reads

\begin{equation}
P_e(q) \sim e^{-\frac{q^2}{2kTC}-\frac{\Delta W q}{KTe}}
		\label{4.11}
\end{equation}

\noindent The equilibrium charge $q_0$ of the condenser is thus

\begin{equation}
q_0 = - \frac{\Delta W \, C}{e}
\label{4.12}
\end{equation}

\noindent These results obtained using equilibrium statistical
mechanics are also found in ref.\cite{vKNL} from the master equation
proposed to describe the system under discussion.  Moreover, the fact
that the condenser in equilibrium bears a charge $q_0$ could be
interpreted as a violation of the second law of thermodynamics.  It is
clear from the fact that the two plates of the condenser are of
different metals that, in equilibrium, no work can be obtained by
connecting them with a wire since the equilibrium voltage $v_0 \equiv
q_0/C = -\Delta W/e$ would exactly be compensated by a contact voltage
barrier due to the difference in electrochemical potentials.

	Consequently, in order to study charge fluctuations in this
system, we have to use eq.(\ref{4.11}) together with eq.(\ref{4.9b}).
In view of fig.2, note that the charge in the condenser decreases when
$I$ is positive.  Thus, the phenomenological equation for this system
is obtained from eq.  (\ref{4.9b}) and reads

\begin{equation}
\dot{q} = -I(v) = I_0 \left(1-  e^{e(q-q_0)/kTC} \right)
\label{4.13}
\end{equation}

\noindent where eq.  (\ref{4.12}) has been used.  Finally, with the
aim of comparing with the results obtained in ref.\cite{vKNL}, we
rewrite eqs.  (\ref{4.11}) and (\ref{4.13}) in terms of the
dimensionless variable $x \equiv -\epsilon^{1/2}(q-q_0)/e$, where
$\epsilon \equiv e^2/kTC$ is a small parameter that is related to the
amplitude of the fluctuations\footnote{This parameter is related to the
inverse of the size of the system, in this case the capacity of the
condenser $C$\cite{vK}. Physically, $\epsilon$ measures the ratio between
the difference in electrostatic energy between two consecutive electron
jumps and $kT$. The jump probability of a second electron is strongly
affected by the jump of the first one if $\epsilon \sim 1$.}.  We get

\begin{equation}
P_e(x) \sim e^{-x^2/2}					\label{4.14}
\end{equation}

\noindent and

\begin{equation}
\dot{x}= j\left(e^{-\epsilon^{1/2}x} -1 \right)       \label{4.15}
\end{equation}

\noindent where $j\equiv I_0 \,\epsilon^{1/2}/e$.  As in the previous
example, from these two equations we obtain the phenomenological
coefficient $\beta(x)$

\begin{equation}
\beta(x) = \frac{j}{x} \left(1-e^{-\epsilon^{1/2}x} \right)
\label{4.16}
\end{equation}

\noindent Finally, making use of eqs.  (\ref{4.14}) and (\ref{4.16})
in eq.  (\ref{4.8c}) we arrive at

\begin{equation}
\frac{\partial}{\partial t} P(x,t) = \frac{\partial}{\partial x}
\frac{j}{x} \left(1-e^{-\epsilon^{1/2}x} \right)  \left[x P(x,t) +
\frac{\partial}{\partial x}
P(x,t) \right]		 \label{4.17}
\end{equation}

	To end this section, we compare eq.  (\ref{4.17}) with the
expansion of the master equation for this model.  We thus expand our
Fokker-Planck equation in powers of $\epsilon^{1/2}$.  We obtain

\begin{eqnarray}
\frac{1}{j\epsilon^{1/2}} \frac{\partial}{\partial t} P(x,t)& =
&\frac{\partial}{\partial x}
\left(x-\frac{1}{2}\epsilon^{1/2}(x^2-1) + \frac{1}{6}\epsilon
(x^3-2x) \right)
P(x,t)
\nonumber \\
&+&\frac{\partial^2}{\partial x^2} \left(1-\frac{1}{2}\epsilon^{1/2}x
+
\frac{1}{6}\epsilon
x^2 \right)  P(x,t) + {\cal O}(\epsilon^{3/2})        \label{4.18}
\end{eqnarray}

\noindent On the other hand, eq.(45) of ref.\cite{vKNL} has the form

\begin{eqnarray}
 \frac{\partial}{\partial \tau} P(x,t)& = &\frac{\partial}{\partial x}
\left(x-\frac{1}{2}\epsilon^{1/2}(x^2-1) + \frac{1}{6}\epsilon
(x^3-3x) \right)
P(x,t)
\nonumber \\
&+&\frac{\partial^2}{\partial x^2} \left(1-\frac{1}{2}\epsilon^{1/2}x
+
\frac{1}{4}\epsilon
(x^2-1) \right)  P(x,t) \nonumber\\
&+ & \frac{\partial^3}{\partial x^3}\left(\frac{1}{6}\epsilon x
\right) P(x,t)
+\frac{1}{12}
\epsilon \frac{\partial^4}{\partial x^4} P(x,t)+ {\cal
O}(\epsilon^{3/2})
\label{4.19}
\end{eqnarray}

\noindent Comparing these last two equations we see that they agree up
to order $\epsilon^{1/2}$ while up to first order, the equation for
the fluctuations is not even a Fokker-Planck equation but higher order
derivatives have appeared.  We then conclude firstly that to the order
of validity of the Fokker-Planck equation itself, ($\epsilon^{1/2}$),
the fluctuations are correctly described by the phenomenological
theory developed along this paper without ambiguity.  Secondly, up to
this order, the coefficients of the Fokker-Planck equation are
functions of the variable $x$, corresponding to a so-called non-linear
Fokker-Planck equation.  The possibility of correctly set non-linear
Fokker-Plank equations from a phenomenological theory is the main
result of this paper.  As we have seen, however, the validity of the
Fokker-Planck equation to describe fluctuations in non-linear systems
is restricted to small fluctuations.

\section{Conclusions}

\setcounter{equation}{0}

	In this paper we have tried to answer the question about
whether it is possible or not to write a Langevin equation to describe
the equilibrium fluctuations of a given macroscopic variable $X$ with
only a phenomenological knowledge of the system.  To this end, we have
proposed the Langevin equation (\ref{1.8b}) containing two terms
accounting for the existence of fluctuations.
In the first place, we have a function
$\tilde{\Gamma}(x)$ which plays the role of a modification of the
phenomenological law at the scale of the fluctuations.  This answers
one of the major criticisms to the use of Langevin equation to
describe fluctuations in non-linear systems\cite{vK}.  In the second place, we
have introduced a causal random force as given in eq. (\ref{1.8}). We have then
analyzed the particular case in which the cumulants of $\tilde{L}(x,t)$ depend
on a single function $\tilde{c}(x)$, which is equivalent to a multiplicative
noise, which is the simplest case of variable-dependent random force. Then, the
function $\tilde{c}(x)$ can be interpreted as an amplitude of the random force
that can vary from point to point.  We have proved that detailed
balance leads, firstly, to the result that $\tilde{L}(x,t)$ must be a Gaussian
process.  One important consequence of this result is that our
Langevin equation (\ref{1.8b}) is equivalent to a Fokker-Planck
equation.  Secondly, we have obtained the relationship between the
functions $\tilde{\Gamma}(x)$ and $\tilde{c}(x)$ given in eq.  (\ref{3.15b}).
While these are general results for even variables, we have seen that they are
only satisfied for odd variables under certain symmetry conditions. In fact, in
this last situation we have
proved that $\tilde{L}(x,t)$ is a Gaussian process if and only if
$\tilde{c}(x)$ is an
even function, which also implies that $\tilde{B}(x)$ must be odd. The symmetry
of this
"renormalized" flux leads in turn to the fact that $\tilde{\Gamma}(x)$ has to
be odd and the Onsager coefficient $\beta(x)$ has to be an even function. This
is an important condition on the phenomenological coefficient that arises from
microscopic reversibility and weak coupling, which has
the same origin as the symmetry of the Onsager coefficients for the crossed
terms in the linear case\cite{dGr}.

        Moreover, we have
seen that the phenomenological knowledge of the system together
with detailed balance are not enough to determine the properties of
the random force. The existence of this ambiguity indicates that,
at this level, different equations for the fluctuations could be
proposed, all satisfying detailed balance and having the
same equilibrium probability distribution, but leading to different dynamics
for the fluctuations\cite{vKNL}. We have then shown that, if the internal
mechanism that causes the random force is weakly coupled with the
variable that fluctuates, as it has clearly been shown in the case of
Brownian motion, detailed balance together with this {\em weak coupling}
assumption suffices to completely determine the Langevin equation.

        In our analysis we have imposed
weak coupling only at a later stage of the discussion
in order to completely specify the properties of the random force, after
having applied detailed balance. It
is important to note, however, that if we had imposed detailed balance
together with weak coupling from the beginning, we would have seen that the
Gaussian nature of $L(x,t)$
follows both for variables even and odd under time-reversal,
that the "renormalized" flux is authomatically acceptable, and that the
symmetry
property of $\beta(x)$ is a natural consequence of both
principles\footnote{This can be proved by treating in eq. (\ref{3.2}) both
$\psi(x)$ and $P_e(x)$ as arbitrary functions and following the same reasoning
as in appendix B but for a general $c(x)$.}.

        The results derived from the previous analysis have been applied to two
non-trivial examples.  To
the range of validity of the form assumed for the random force in eq.
(\ref{1.8}),
the results coincide with more detailed models for the dynamics of the
fluctuations in those systems.  Taking into account that for
a general process the
validity of the Fokker-Planck equation as an approximation of the
master equation is restricted to small fluctuations\cite{vK},  in the second
example of section 4 we have seen that, to the range of validity of the
Fokker-Planck equation itself, our approach is correct and furnishes
a description of the fluctuations in a genuinely non-linear system.  Therefore,
all the results found here by means of a phenomenological analysis
are consistent with those obtained from the expansion of the master
equation.

	Let us analyze the case $\beta$ constant in more detail. From
eqs. (\ref{4.7b}) and (\ref{4.8b}), we get that
$\tilde{c}^2=\beta$ (also constant) and $\tilde{\Gamma}=0$, in agreement with
ref.\cite{Maz1} which, for linear Langevin equations, reduces to the standard
results. It is important to realize that detailed balance
alone ensures, in view of eq. (\ref{3.15b}), that $<\tilde{\Gamma}(x)>=0$, but
still $\tilde{\Gamma}(x)$ and $\tilde{c}(x)$ can be non-constant functions
related by eq. (\ref{3.15b}). Therefore, the standard theory for
Langevin equations, either linear or with constant $\beta$, relies on an
implicit
weak coupling assumption\footnote{This implicit weak coupling assumption is
formulated by demanding that the random force $F(x(t),t)$ is
independent of the variable $x(t)$.}.

On the basis of causality and stationarity, the second
moment of the random force in a non-linear Langevin equation of the type
(\ref{2.1}), has been proved to satisfy\cite{Maz2}

\begin{equation}
\langle F(x(t),t) F(x(t'),t')\rangle  = \langle \tilde{L}(x(t),t)
\tilde{L}(x(t'),t')\rangle
= 2 \langle x(t) \tilde{B}(x(t))\rangle  \, \delta(t-t')       \label{5.0b}
\end{equation}

\noindent Inserting the definition of $\tilde{B}(x)$ in the right hand side of
this equation, after partial integration we get

\begin{equation}
\langle x(t) \tilde{B}(x(t))\rangle  = \langle \beta(x(t))\rangle +\left\langle
x(t) \left. \frac{d}{dx} \beta(x)\right|_{x=x(t)}\right\rangle  -\langle x(t)
\tilde{\Gamma}(x(t))\rangle           \label{5.0c}
\end{equation}

\noindent If detailed balance applies, one can
make use of eq. (\ref{4.8b}). Then, the second and third terms in
this last equation cancel and we finally obtain\cite{Maz3}

\begin{equation}
\langle F(x(t),t) F(x(t'),t')\rangle  =  2 \langle \beta(x(t))\rangle  \,
\delta(t-t')                 \label{5.0e}
\end{equation}

\noindent Again, if $\beta$ is constant, only now we recover the
fluctuation-dissipation theorem as a consequence of both detailed balance and
weak coupling. In the genuinely non-linear case,
eq. (\ref{5.0e}) as well as eq.(\ref{4.7b}) are reminiscent of the
fluctuation-dissipation theorem existing for linear systems.
In particular, eq. (\ref{4.7b}) relates the amplitude of the random force,
$\tilde{c}(x)$, to the
phenomenological coefficient $\beta (x)$ at every point.

        Finally, one can
calculate the average of the flux in equilibrium from eq.
(\ref{1.8b}). Effectively, from causality and eq. (\ref{3.15b}) it follows that

\begin{equation}
\langle \dot{x}(t)\rangle = \left\langle \beta(x(t)) \left. \frac{d}{dx} \ln
P_e(x)\right|_{x=x(t)}+\left.\underline{\frac{d}{dx} \beta(x)}\right|_{x=x(t)}
\right\rangle \equiv 0          \label{5.2}
\end{equation}

\noindent as expected.  Notice that the underlined term is the
modification of the phenomenological law due to the fluctuations and
ensures that $<\dot{x}>$ is zero in equilibrium, so that there is no
violation of the second law as it is the case in the so-called
"Brillouin paradox"\cite{vK}.

\section*{Acknowledgements}

        The authors whish to warmly thank P. Mazur for helpful discussions and
encouragement and to D. Bedeaux and J. Bafaluy for his support.

\appendix

\renewcommand{\theequation}{\Alph{section}.\arabic{equation}}

\section{Derivation of the master equation}
\setcounter{equation}{0}

        The density distribution given in eq. (\ref{2.3}) satisfies
a continuity equation

\begin{equation}
\frac{\partial}{\partial t} \rho(x,t) = -\frac{\partial}{\partial x}
\dot{x}(x,t)
\, \rho(x,t)                                  \label{2.6}
\end{equation}

\noindent In view of eq.  (\ref{2.3}), and due to the properties of
the $\delta$-function, to compute $\dot{x}(x,t)$ we can use eq.
(\ref{2.2d}) replacing the random variable $x(t)$ by the field variable
$x$, giving

\begin{equation}
\frac{\partial}{\partial t} \rho(x,t) =  -\frac{\partial}{\partial x}
\left(-B(x)+L(x,t)  \right) \rho(x,t)            \label{2.7}
\end{equation}

\noindent where use has been made of eq. (\ref{1.8}). Eq. (\ref{2.7}) can be
formally solved, obtaining

\begin{equation}
\rho(x,t) = \left\{ \exp \int_0^t dt' \, \frac{\partial}{\partial x}
\left( B(x)
-L(x,t) \right) \right\} \delta(x-x_0)                \label{2.8}
\end{equation}

\noindent where use has been made of $\rho(x,t=0) = \delta(x-x_0)$.
Note that the term between curly brackets is independent of the
initial condition $x_0$ since $x$ is not a random variable but a field
point and the process $L(x,t)$ has been assumed to be independent of
$x(t)$, so that it is also independent of $x(t=0)=x_0$.  Thus, we can
average both members of this last equation and use eq.  (\ref{2.4}) to
also derive the formal solution for the conditional probability
density

\begin{equation}
P(x,t|x_0) = \left\langle \exp \int_0^t dt' \,
\frac{\partial}{\partial x}
\left( B(x)
-L(x,t) \right) \right\rangle P(x,t=0|x_0)          \label{2.9}
\end{equation}

\noindent The average is performed over all the realizations of $L(x,t)$
and it is independent of $x_0$.  This last expression can also be
written in terms of the cumulants

\begin{equation}
P(x,t|x_0) = \exp \left\{ \sum_{n=1}^{\infty} \frac{1}{n!}
\left\langle \left(\int_0^t dt' \, \frac{\partial}{\partial x} \left[
B(x)
-L(x,t) \right] \right)^n \right\rangle_c \right\} P(x,t=0|x_0)
\label{2.10}
\end{equation}

\noindent For $n=1$ one has

\begin{equation}
\left\langle \left(\int_0^t dt' \, \frac{\partial}{\partial x} \left[
B(x)
-L(x,t) \right] \right) \right\rangle_c = \int_0^t dt' \,
\frac{\partial}{\partial
x} \left[ B(x) -<L(x,t)>_c \right]
= t \frac{\partial}{\partial x} B(x)       \label{2.11}
\end{equation}

\noindent For $n \geq 2$ one has

\begin{equation}
\left\langle \left(\int_0^t dt' \, \frac{\partial}{\partial
x} \left[
B(x)
-L(x,t) \right] \right)^n \right\rangle_c
= (-1)^n \, t \, \alpha_n \left( \frac{\partial}{\partial x} c(x)
\right)^n
                \label{2.12}
\end{equation}

\noindent where in deriving the last line use has been made of
eqs.(\ref{2.2c}) and (\ref{nova1}). Therefore, substituting these results into
eq. (\ref{2.10}) one gets

\begin{equation}
P(x,t|x_0) = \exp \left\{ t \left[\frac{\partial}{\partial x} B(x) +
\sum_{n=2}^{\infty} (-1)^n \frac{1}{n!} \alpha_n \left(
\frac{\partial}{\partial x} c(x) \right)^n \right]\right\}
P(x,t=0|x_0)
                                        \label{2.13}
\end{equation}

\noindent Differentiating with respect to time, one arrives at eq.
(\ref{2.14}).

\section{Proof of $\alpha_i=0$ for $i\geq 3$.}
\setcounter{equation}{0}

	In this appendix we will show the possible solutions of the
system of equations (\ref{3.10}) and (\ref{3.20}) for the parameters
$\alpha_i$, $i\geq 2$. Once removed $\alpha_2$ (recall that the coefficient of
the only term involving this constant is zero), the
system is a
homogeneous set of equations for $\alpha_i$ for $i \geq 3$ with $x$-dependent
coefficients. Its solution is
$\alpha_i=0$ for $i\geq 3$ if the determinant is non-zero and, therefore,
$L(x,t)$ is a Gaussian process.  However, if its determinant is
zero, then other solutions with some $\alpha_i\neq 0$ may exist.  In
order to find the expression of the determinant, let us separately
write the equations for $m$ even and odd. From (\ref{3.10})
and (\ref{3.20}) it follows that

\begin{eqnarray}
-\frac{1}{(2 k)!}{\cal D}y(x) \alpha_{2 k+1}+\sum_{n=2
k+2}^{\infty}
\frac{1}{n!}\left(
\begin{array}{c} n \\ 2 k \end{array} \right) {\cal D}^{n-2
k}y(x)\alpha_{n}&=&0;	\;\;\;\;\;\; m=2 k
\label{b.1}\\
-\frac{1+\epsilon}{(2 k+1)!}y(x) \alpha_{2 k+1}+\sum_{n=2 k+2}^{\infty}
\frac{1}{n!}\left( \begin{array}{c} n \\ 2 k+1 \end{array} \right)
{\cal
D}^{n-(2
k+1)}y(x)\alpha_{n}&=&0;		\;\;\;\;\;\; m=2 k+1
				 \label{b.2}
\end{eqnarray}

\noindent with $k=1,2,...$.  We have defined the function $y(x) \equiv c(x)
P_e(x)$ for the ease of notation.

        Let us first consider the case of even variables. We thus replace
$\epsilon$ by $1$ in eq. (\ref{b.2}). Making now use of this last equation,
we can express the coefficient of $\alpha_{2 k+1}$ in terms of all higher ones.
If one then proceeds to substitute eqs.(\ref{b.2}) systematically for all $k$
in eqs.  (\ref{b.1}), one arrives at a set of equations only for the even
coefficients.  Moreover, this new set of equations is triangular and
it is then sufficient to calculate the terms of the diagonal in order
to know its determinant.  One can see that the determinant is zero only if

\begin{equation}
y(x) \left(-({\cal D} y(x))^2+y(x) {\cal D}^2 y(x) \right) =0
					\label{b.3}
\end{equation}

\noindent which implies that some $\alpha_i$ may be non-zero if $y(x)$
satisfies the non-linear differential equation

\begin{equation}
-c^2(x) \left( \frac{d y}{d x} \right)^2+c(x) y(x) \frac{d c}{d x}
\frac{d y}{d
x}+c^2(x)
y(x) \frac{d^2 y}{d x^2}=0
							\label{b.4}
 \end{equation}

\noindent whose solutions are

\begin{equation}
y(x)=c(x) P_e(x)=\nu+\mu \int P_e(x) d x
					\label{b.5}
\end{equation}

\noindent $\nu$ and $\mu$ being integration constants. This
last equation determines $c(x)$, since $P_e(x)$ has been
considered as given. Note that in this particular case, $c(x)$ is
independent of the phenomenological coefficient $\beta
(x)$. Furthermore, these solutions are eigenfunctions of the
differential operator
${\cal D}$,

\begin{equation}
{\cal D}y(x)=\mu y(x)
			\label{b.6}
\end{equation}

\noindent In order to determine the set of cumulants $\alpha_i$, we
will use this last property in eqs. (\ref{3.8}) and (\ref{3.9}). We
get

\begin{eqnarray}
\frac{{\cal D}B(x) P_e
(x)}{y(x)}&=&-\sum_{n=2}^{\infty}\frac{(-1)^n}{n!}\mu^n
\alpha_n 		\label{b.7}\\
\frac{2 B(x) P_e (x)}{y(x)}&=&-\sum_{n=2}^{\infty}\frac{(-1)^n
n}{n!}\mu^{n-1}
\alpha_n		\label{b.8}
\end{eqnarray}

Let us introduce the function

\begin{equation}
\psi(\mu)=\sum_{n=2}^{\infty}\frac{(-1)^n}{n!}\mu^n \alpha_n
		\label{b.9}
\end{equation}

\noindent in terms of which equations (\ref{b.7}) and (\ref{b.8})
become

\begin{eqnarray}
\frac{{\cal D}B(x) P_e (x)}{y(x)}&=&-\psi(\mu)    \label{b.10}\\
\frac{2 B(x) P_e (x)}{y(x)}&=&-\frac{\partial \psi(\mu)}{\partial \mu}
                      \label{b.11}
\end{eqnarray}

\noindent Thus, eliminating $B(x)$ between eqs. (\ref{b.10}) and (\ref{b.11}),
we obtain a differential equation for $\psi (\mu)$

\begin{equation}
\frac{d \psi}{d \mu}-\frac{ 2 \psi}{\mu}=0
	\label{b.13}
\end{equation}

\noindent which indicates that $\psi$ is a quadratic function of the
parameter $\mu$.  This implies that all the cumulants of $L(x,t)$ should
be zero except for the second one.  Therefore, we have proved that
also in this case $L(x,t)$ is Gaussian. Moreover, from eqs.
(\ref{b.9}-\ref{b.11}) we arrive at the corresponding Langevin equation

\begin{equation}
\frac{d x}{dt}=\frac{c(x)}{2} \alpha_2 \mu+L(x,t)
\label{b.14}
\end{equation}

\noindent Now, in view of eq. (\ref{b.5}), we see that eq. (\ref{b.14}) is
independent of the phenomenological coefficient $\beta(x)$,
corresponding to a situation which is out of the scope of
this work.

	Let us finally point out that the special case $\mu=0$ leads
to a Langevin equation of the form

\begin{equation}
\frac{d x}{dt}=L(x,t)				\label{b.15}
\end{equation}

\noindent which is again non-acceptable.  In this case,
however, detailed balance does not suffices to completely fix the properties of
$L(x,t)$.  Although all the odd cumulants are zero, the even
ones remain undetermined, allowing therefore for non-Gaussian noises.

        For odd variables, we have to replace $\epsilon$ by $-1$ in eqs.
(\ref{b.1}) and (\ref{b.2}). In this case, the first term in eq. (\ref{b.2})
vanishes. Therefore, the homogeneous system of equations is already
triangular, showing that $\alpha_i =0$ for $i \geq 3$, unless the
determinant is zero. This happens only if

\begin{equation}
{\cal D} y = 0          \label{b.18}
\end{equation}

\noindent In this case, nothing can be said about the set of cumulants
$\alpha_i$ since in view of eq. (\ref{b.18}), in eqs. (\ref{b.1}) and
(\ref{b.2}) all the coefficients vanish. The corresponding Langevin equation
also has the form given in eq. (\ref{b.15}), again non-acceptable.

\newpage

\section*{Figure Captions}

\noindent Fig.1:  Alkemade's diode.  Thick lines represent plates and
wires of metal $1$ while thin lines represent plates and wires of
metal $2$.  The system enclosed by the frame corresponds to the diode
and it is kept at constant temperature $T$.  The figure on the right
hand side represents a schematic view of the system.  Note that the
junction of metals $1$ and $2$ give rise to a potential barrier.

\noindent Fig.  2:  Alkemade's diode in parallel with a condenser.  To
avoid the junction in the wires, the plates of the condenser are of
different metals.
We study the charge fluctuations $q$ in the condenser of capacity $C$.


\begin{thebibliography}{99}

\bibitem{Lang} P. Langevin, Comptes Rendus Acad. Sci. (Paris) {\bf
146}, 530 (1908).

\bibitem{Ors} G.E. Uhlenbeck and L.S. Ornstein, Phys. Rev. {\bf
34}, 823 (1930), reprinted in N. Wax, {\em Selected Papers on Noise ans
Stochastic Process} (Dover, New York, 1954).

\bibitem{vK} N.G. van Kampen, {Stochastic Processes in Physics and
Chemistry}, (North Holland, Amsterdam, 1992) 2nd. edition.

\bibitem{Lan} L. Landau and E.M. Lifshitz, {\em Fluid Mechanics}
(Pergamon, New York, 1959).

\bibitem{Ons} L. Onsager and S. Machlup, Phys. Rev. {\bf 36}, 823 (1953).

\bibitem{dGr} S.R. de Groot and P. Mazur, {\em Non-equilibrium
Thermodynamics} (Dover, New York, 1984).

\bibitem{vK2} See ref.\cite{vK}, chap. IX, \S 4-5; in particular pags.
228 and 234-237.

\bibitem{vK3} See ref.\cite{vK}, p. 235.

\bibitem{Maz1} P. Mazur and D. Bedeaux, Physica A {\bf 173}, 155 (1991);
P. Mazur and D. Bedeaux, Biophysical Chemistry {\bf 41}, 41 (1991).

\bibitem{Maz2} P. Mazur, Phys. Rev. A {\bf 45}, 8957 (1992).

\bibitem{Maz2b} P. Mazur and D. Bedeaux, Langmuir {\bf 8}, 2947 (1992).

\bibitem{Alk} C.T.J. Alkemade, Physica {\bf 24}, 1029 (1958).

\bibitem{vKNL} N.G. van Kampen in {\em Fluctuation Phenomena in
Solids}, R.E. Burgess ed. (Academic Press, New York 1965), Ch. 5.

\bibitem{dGr2} See ref.\cite{dGr}, p. 98-99.; See also ref.\cite{vK},
chap. V \S 4 and 6 and references therein.

\bibitem{Sa} J.M. Sancho, M. San Miguel and D. D\"{u}rr, J. Stat.
Phys. {\bf 28}, 291 (1982).

\bibitem{Mun} A. Munster {\em Classical Thermodynamics} John Wiley
\& Sons (London, 1970).

\bibitem{Maz3} P. Mazur (private comunication).

\end{thebibliography}
\end{document}